# Diagnostic performance of radiologists with and without different CAD systems for mammography

Adele Lauria[*a], M. Evelina Fantacci[b], Ubaldo Bottigli[c], Pasquale Delogu[b], Francesco Fauci[d], Bruno Golosio[c], P.L. Indovina[a], G. Luca Masala[c], Piernicola Oliva[c], Rosa Palmiero[a], Giuseppe Raso[d], Simone Stumbo[c], Sabina Tangaro[b]
[a]University of Napoli "Federico II", Italy
[b]INFN and University of Pisa, Italy
[c]INFN of Cagliari and University of Sassari, Italy
[d]University of Palermo, Italy

## ABSTRACT

The purpose of this study is the evaluation of the variation of performance in terms of sensitivity and specificity of two radiologists with different experience in mammography, with and without the assistance of two different CAD systems. The CAD considered are SecondLook[TM] (CADx Medical Systems, Canada), and CALMA (Computer Assisted Library in MAmmography). The first is a commercial system, the other is the result of a a research project, supported by INFN (Istituto Nazionale di Fisica Nucleare, Italy); their characteristics have been already reported in literature. To compare the results with and without these tools, a dataset composed by 70 images of patients with cancer (biopsy proven) and 120 images of healthy breasts (with a three years follow up) has been collected. All the images have been digitized and analysed by two CAD, then two radiologists with respectively 6 and 2 years of experience in mammography indipendently made their diagnosis without and with, the support of the two CAD systems. In this work sensitivity and specificity variation, the $A_z$ area under the ROC curve, are reported. The results show that the use of a CAD allows for a substantial increment in sensitivity and a less pronounced decrement in specificity. The extent of these effects depends on the experience of the readers and is comparable for the two CAD considered.

## 1. INTRODUCTION

In this study two CAD systems have been used as second reader. The advantages and disvantage that their use produced on the radiologist's work are reported.
It was demonstrated that the use of computerized detection in mammography produced an increment in sensitivity, both for masses and for microcalcification clusters research [1,2,3,4,5]. In medical physics literature the increasing of sensitivity in finding spiculated lesions ranges from 3% to 14%. The automatic classification of microcalcifications by neural networks shows a good performance to distinguish benign from malignant clusters [6]. Computerized systems are able to find the signs of cancer lost by a human double reading [7].
In literature few studies have compared two different CAD systems using the same data-set of images [8].
In this work the evaluation of sensitivity and specificity, in the detection of microcalcification clusters, of 2 radiologists with and without the help of two different CAD systems is reported. The performances of the two CAD systems working alone have been already evaluated in previous works [9,10].
As well a qualitative study on the new way to make a medical report with the aid of computer is realized as well. The readers viewed the image on different supports: first the original image has been viewed on film, then the same image, with the area suggested from the CAD system, was visualized on paper (SecondLook) or on monitor (CALMA). We investigated how the double reading with the CAD influences radiologists in their work-flow, in terms of time and loss of concentrantion.

---

[*]adele.lauria@na.infn.it; phone +39081676121; fax +39081676346; http://na.infn.it; INFN and University of Napoli "Federico II", Complesso Universitario Monte S. Angelo, Via Cinthia, I-80126, Napoli, Italy.
.

## 2. CAD SYSTEMS

### 2.1 SecondLook[TM] CAD

The first CAD system considered was the SecondLook[TM] (CADx, Medical Systems) produced in Canada. It runs in three steps. First it digitilizes mammograms, then a neural network analizes them to produce, as last step, a printed output (called Mammagraph) where the regions of interest are pointed out by markers. An oval pointer indicates a massive lesion, while a rectangular one points out a cluster of microcalcification. Time to obtain the print out is about 6 minutes. It was not possible to modify, to visualize or to store images. The print out is not useful to realize a diagnosis because the low quality [11]. The SecondLook printout is shown in figure 1.
Preliminary results about its perfomances have been published elsewhere [12].

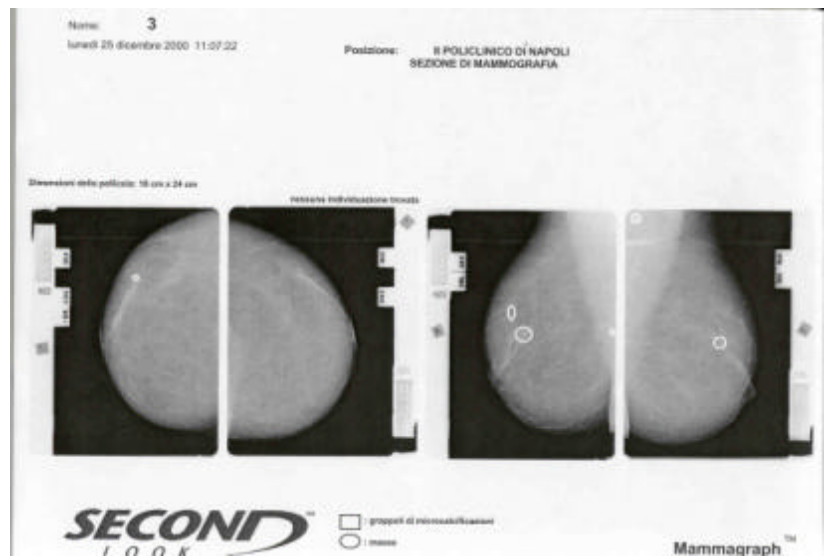

Fig.1: SecondLook printout

### 2.2 CALMA CAD

The other CAD system used is CALMA (Computer Assisted Library in MAmmography), developed by INFN (*Istituto Nazionale di Fisica Nucleare*), some universities and many Italian hospitals. As a first step the mammogram is digitized by a CCD linear scanner (85 ?m, 12 bit/pixel), and it is stored in a CALMA specific format. In the following step, the radiologist visualizes it on the screen with help of some visualization tools (i.d. zoom, contrast enhancement). Microcalcification icon starts the program automatic search. This program contains both algorithms and neural networks. In few seconds a red rectangle points out regions of interest (ROI) where suspicious microcalcification clusters are recognized by the system (Figure 2). Software and hardware characteristics of the CALMA system can be found in literature [13,14,15].

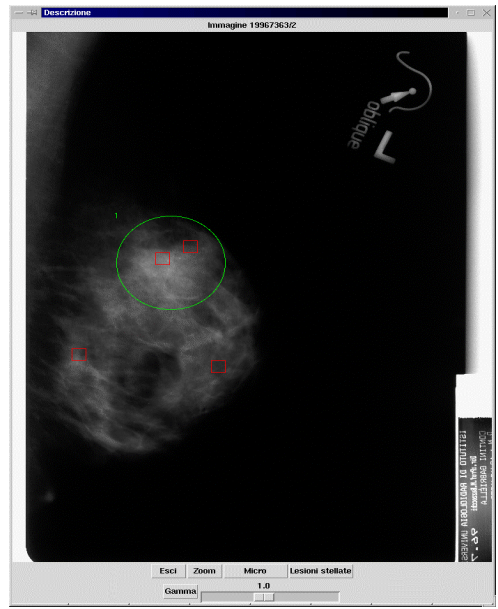

Fig. 2: The CALMA clusters microcalcification reserch

Both with CADx and CALMA require 2.5 minutes to digitalize and visualize ROIs on images. The ROC curve plotted for CALMA CAD is reported in Figure 3.

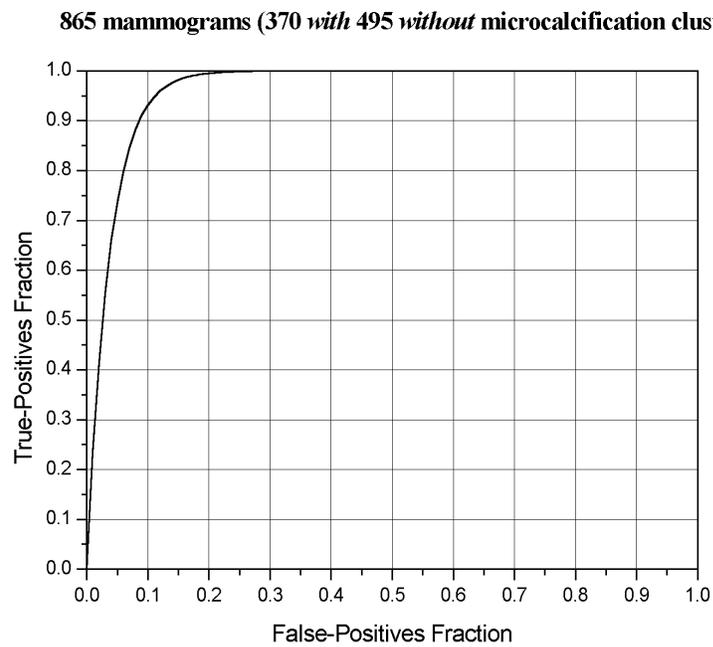

Fig. 3: ROC curve of CALMA

The dataset used is composed by 70 images of malignant histopathological patients (biopsy proven) and 120 images of healthy cases (proven by three years radiological follow up). The images were collected from different mammographic centers, from screening and clinical tests. In the detected images different kind of microcalcification clusters are present. Among them there were 24 cases of ductal infiltranting carcinoma, 6 cases of ductal in situ carcinoma and 2 cases of tubular infiltranting carcinoma, in the the other cases the kind of carcinoma was not specified. Each cluster had a diameter ranging from 3 mm to 4 cm.

All the images weredigitized and analysed by the two CAD systems, then two radiologists with respectively 6 and 2 years of experience in mammography in a public hospital viewed the films in a conventional way at the diaphanoscope. Each radiologist viewed all the images in a random order, every time he was blind to the final diagnosis and independent from the other one. After two months the radiologists viewed the same films with the SecondLook printout; 10 months later they examined the digitized images on a high resolution screen, observing the ROI suggested by CALMA CAD. The medical report was compared to the final diagnosis. The goal consisted of the detection of the right area where the lesion was biopsy proven.

Time to take decision and remarks of radiologists concerning the new procedures are reported. All the sessions were repeated in the same physical conditions of light, time and no hurry. The doctors were previously informed about how the systems work.

Different parameters were considered to evaluate the sigificativity of the results from a diagnostic and statistic point of view. For each radiologist sensitivity and specificity were estimated: a) alone, b) with CADx, c) with CALMA, according to the following relations:

Sensitivity = *number of images properly classified as positive (TP) / total number of positive images (TP+FN);*

Specificity = *number of images properly classified as negative (TN) / total number of negative images (TN+FP).*

TP (true positive) or TN (true negative) are obtained when the medical report is in accord with the final diagnosis, while FP (false positive) or FN (false negative) indicate that the human diagnosis doesn't match with the gold standard. Sensitivity and specificity are not sufficient to describe the efficacy of a CAD system [16]. Another parameter is commonly considered: the area $A_z$ under the ROC curve. The ROC curve is the most complete way to describe the diagnostic capability of a system, it represents the capability to distinguish positive from negative cases. The yes/no method [17] was employed to calculate it. Sensitivity, specificity and ROC analysis were evaluated using the bootstrap method, that estimates the 95% confidence intervals (CIs) in the statistical analisys. Matlab software was used to perform this calculation. Using chi-square test, *p* values were evaluated. The area $A_z$ under the ROC curve and *p* values were calculated using MedCalc? software.

## 4. RESULTS

Radiologist A detected 58 (58/70) images of cancer cases and 110 (110/120) images of healty cases; radiologist B detected 50 (50/70) images of cancer cases and 89 (89/120) images of healty cases.

Using the CADx system as second reader radiologist A detected 66 (66/70) images of cancer cases and 101 (101/120) images of healty cases; radiologist B detected 58 (58/60) images of cancer cases and 85 (85/120) images of healty cases.

Using the CALMA system as second reader radiologist A detected 66 (66/70) images of cancer cases and 106 (106/120) images of healty cases; radiologist B detected 61 (61/70) images of cancer cases and 85 (85/120) images of healty cases.

Sensitivity and specificity, in percentage, are respectively reported in table 1 and in table 2. These values are estimated to 95% confidence interval (C.I.).

Using chi-square test *p* values are evaluated; they always result less than 0.0001, only in one case (reader A alone) *p* is equal to 0.0005.

|   | Alone (C.I.) | With CADx (C.I.) | With CALMA (C.I.) |
|---|---|---|---|
| A | 82.8% (4.5%) | 94.3% (2.8%) | 94.3% (2.8%) |
| B | 71.5% [1] (5.4%) | 82.9% (4.5%) | 87.1% (4.0%) |

Tab. 1: Sensitivity ($p<0.0001$) [1]$p=0.0005$

|   | Alone (C.I.) | With CADx (C.I.) | With CALMA (C.I.) |
|---|---|---|---|
| A | 87.5% (3.0%) | 84.2% (3.3%) | 87.5% (3.0%) |
| B | 74.2% (4.0%) | 70.8% (4.2%) | 70.9% (4.1%) |

Tab. 2: Specificity ($p<0.0001$)

|   | Alone | With CADx | With CALMA |
|---|---|---|---|
| A | 0.85 | 0.89 | 0.91 |
| B | 0.73 | 0.77 | 0.79 |

Tab. 3: $A_z$ values

## 5. CONCLUSION

The average value of sensitivity for a radiologist alone is 78.1% ?5.9%, the average value of sensitivity with the aid of SecondLook and CALMA CAD is respectively 88.5% ? 5.7% and 90.5% ?3.6%. These values, are in accord with previous works [4, 5, 6, 7], where an increment in sensitivity is shown when a CAD system is used as second reader. In addition the standard deviation decreases by a small amount. In a recent studiy Jiang and et al. [18] reported the value of sensitivity of 74% ? 11% of radiologists working alone; using the R2 technology CAD system, the value improved by 13% becoming 87% ? 9%. These results are in line with ours where the average improvement in sensitivity for CADx and for CALMA is respectively 10.4% and 12.4%. Taft et al. [10] also reported an increment in sensitivity with the aid of CADx ranging from 8% to 10.5%. It is interesting to notice that the average decrease in specificity of the present study is 4.2% for CADX and 2.2% for CALMA. The area $A_z$ under the ROC curve increased in both studies but the improvements evaluated in the present work are not statistically significant.

The results reported here are also in accord with studies performed by Ciatto and colleagues [19], and by Malich and colleagues [12].

The diagnosis method of the radiologists involved evolved in the course of this study. The reading time, at the beginning, using either of the CAD systems was twice as long, but during the study the radiologists learned to synchronize the double reading.

In conclusion, previous studies have shown that the double reading produces an increment in terms of sensitivity of up to 15% [2, 3]. In our study, not conditioned by the dataset, we show that the CAD system as a second reader detemined an increase in overall sensitivity of up to 15.6%, with a slight increase of the number of false positives. The decrease in specificity was more significant for the least experienced of the radiologists. $A_z$ area increased both with CADx and CALMA, the variation ranges from 0.01 to 0.06, independently of the skill of reader. The results obtained show that CAD systems could be advantegeously used in clinical practice specially when the human double reading is not feasible.